# A Big Data Driven Framework for Duplicate Device Detection from Multi-sourced Mobile Device Location Data

**Aliakbar kabiri (**email**:** kabiri@umd.edu) (ORCID: 0000-0003-2119-007X)
(Corresponding Author)
Graduate Research Assistant
Department of Civil and Environmental Engineering
University of Maryland, 1173 Glenn Martin Hall, College Park, MD 20742, United States

**Aref Darzi (**email**:** adarzi@umd.edu) (ORCID: 0000-0003-2558-5570)
Faculty Assistant
Department of Civil and Environmental Engineering
University of Maryland, 1173 Glenn Martin Hall, College Park, MD 20742, United States

**Saeed Saleh Namadi (**email**:** saeed@umd.edu) (ORCID: 0000-0002-0182-1543)
Graduate Research Assistant
Department of Civil and Environmental Engineering
University of Maryland, 1173 Glenn Martin Hall, College Park, MD 20742, United States

**Yixuan Pan (**email**:** ypan1003@umd.edu) (ORCID: 0000-0002-5015-0088)
Research Scientist
Department of Civil and Environmental Engineering
University of Maryland, 1173 Glenn Martin Hall, College Park, MD 20742, United States

**Guangchen Zhao (**email**:** gczhao@umd.edu) (ORCID: 0000-0002-0803-207X)
Graduate Research Assistant
Department of Civil and Environmental Engineering
University of Maryland, 1173 Glenn Martin Hall, College Park, MD 20742, United States

**Qianqian Sun (**email**:** qsun12@umd.edu) (ORCID: 0000-0003-3684-4603)
Graduate Research Assistant
Department of Civil and Environmental Engineering
University of Maryland, 1173 Glenn Martin Hall, College Park, MD 20742, United States

**Mofeng Yang (**email**:** mofeng@umd.edu) (ORCID: 0000-0002-0525-7978)
Ph.D. Candidate
Department of Civil and Environmental Engineering
University of Maryland, 1173 Glenn Martin Hall, College Park, MD 20742, United States

**Mohammad Ashoori (**email**:** mashoori@umd.edu**)** (ORCID: 0000-0001-7188-2053)
Graduate Research Assistant
Department of Civil and Environmental Engineering
University of Maryland, 1173 Glenn Martin Hall, College Park, MD 20742, United States

**ABSTRACT**
Mobile Device Location Data (MDLD) has been popularly utilized in various fields. Yet its large-scale applications are limited because of either biased or insufficient spatial coverage of the data from individual data vendors. One approach to improve the data coverage is to leverage the data from multiple data vendors and integrate them to build a more representative dataset. For data integration, further treatments on the multi-sourced dataset are required due to several reasons. First, the possibility of carrying more than one device could result in duplicated observations from the same data subject. Additionally, when utilizing multiple data sources, the same device might be captured by more than one data provider. Our paper proposes a data integration methodology for multi-sourced data to investigate the feasibility of integrating data from several sources without introducing additional biases to the data. By leveraging the uniqueness of travel pattern of each device, duplicate devices are identified. The proposed methodology is shown to be cost-effective while it achieves the desired accuracy level. Our findings suggest that devices sharing the same imputed home location and the top five most-visited locations during a month can represent the same user in the MDLD. It is shown that more than 99.6% of the sample devices having the aforementioned attribute in common are observed at the same location simultaneously. Finally, the proposed algorithm has been successfully applied to the national-level MDLD of 2020 to produce the national passenger origin-destination data for the NextGeneration National Household Travel Survey (NextGen NHTS) program.

## INTRODUCTION

A key aspect of transportation planning is understanding the population's travel behavior. For many years, travel surveys are one of the most reliable methods to obtain the movement patterns of a population. The National Household Travel Survey (NHTS) and the Maryland Statewide Travel Survey (MTS) are just two examples of travel surveys that agencies conducted to collect the travel diaries of a sample of the residents to be used in studies and make proper decisions in the transportation field. Although travel surveys provide many insights into the population's movement patterns, some drawbacks are associated with them. For large-scale studies, it is often impossible to have a high sampling rate and a long study period. Accordingly, in both surveys, a small population sample is surveyed for a short period of time (i.e., a couple of days), and then the observed patterns are expanded to the entire population. This could lead to biases in several aspects, such as different demographic characteristics of population groups, temporal bias. For this reason, modern data collection and processing methods should be used in such studies for more accurate disclosure of travel patterns and to provide a larger sample size and a more extended study period.

As technology has grown and mobile phones have become ubiquitous, a vast portion of the population has access to devices with Global Positioning Systems (GPS). Nowadays, many mobile phone applications collect users' locations in latitude-longitude and timestamps showing the time the location was reported. This offers a great source of information about the travel patterns of the population. However, when it comes to utilizing these datasets, several data processing and algorithms are required to explore the underlying travel behaviors of the people. This study aims to utilize Mobile Device Location Data (MDLD) for such analysis and data processing.

MDLD is collected and gathered by several data vendors using different applications installed on smart devices. Therefore, a user's location might be reported by more than one application on a device. Furthermore, in cases where a sample with a high penetration rate and frequency is needed, multiple data vendors' data could be integrated to achieve a higher data quality. However, this step may lead to observe the same user multiple times in the data with different device identifiers. Moreover, electronic devices are more widely available nowadays. As a result, same data subject might have two or more devices when traveling, such as a phone, a tablet, and a computer. As a result, multiple trajectories may be generated with different device identifiers (IDs) for the same user. These three reasons highlight the need for device deduplication methodology to identify such problematic cases in the data in order to avoid introducing additional biases. In this paper, a novel deduplication algorithm is developed to avoid overrepresenting a user's travel patterns in the data sample.

In the following sections, first, a brief literature review on MDLD and device deduplication methods are discussed; then, the dataset used for this study is described and the methodology used for this analysis is explained; afterward, findings and validations are illustrated; and finally, conclusions and future directions are discussed.

**LITERATURE REVIEW**

In recent years, mobile device location data have become popular for studying the travel behavior of the populations. These datasets mainly include the records gathered from call detail records (CDRs), sightings data, GPS-based technology data, or location-based service data. The following are brief descriptions of each of these valuable data sources.

In-vehicle GPS technology reports the location of the vehicles every few seconds. Much research incorporates this kind of data into their analyses. Chankaew et al. (2018) analyzed freight traffic using national truck GPS data in Thailand [1]. CDRs are records that are produced by a telephone exchange or any other telecommunications equipment. These records contain the callers' phone numbers, starting time of the call, duration, and other phone call information. These data report the location of the cell towers instead of a user's actual location [2]. On the other hand, a less frequently used dataset, called sightings, is generated each time the phone is located. What sightings data report are the location of the device using triangulation of multiple towers [2]. Finally, Location-based Service (LBS) data consist of location information recorded by smartphone applications using GPS, cellular towers, Wi-Fi, and other types of connections to track the device's location [3]. This study employs LBS data.

Before utilizing mobile device location data, several data preprocessing steps need to be done and one of these steps is to determine whether all the device identifiers belong to unique individuals, or some individuals are represented by multiple identifiers due to the reasons discussed in the introduction section. Studies on the uniqueness of the data subjects, i.e., the population, can be classified into two categories: research on the feasibility of using demographic information and the feasibility of using spatiotemporal information. Among studies that evaluated the demographic information to be used in device uniqueness identification, a study on 1990 census data by Sweeney (2000) showed that 87% of the U.S. population could be identified uniquely using the collection of demographic attributes such as gender, date of birth, and 5-digit zip code [4]. Furthermore, nearly 50% of the population can be uniquely identified by having their place of residency, gender, and date of birth. Golle (2006) used the 2000 census data to revisit the uniqueness of individuals using the same demographic information and revealed that the previous ratio decreased from 87% to 63% [5].

Other studies that used the spatiotemporal information of mobile device location data and are more pertinent to this paper include research by Trestian et al. (2009). They noted that people spend most of their time in their comfort zone, defined as the top three visited locations [6]. Based on this study, those who stayed in five base stations during a week spent about 90% of their time in the top three locations. Even when a user had 50 base stations defined as areas visited having the size of on average 4 square kilometers, about 55% of their time was spent in their top three visited locations; this indicates that devices representing the same individuals are likely to share the top three visited regions. Golle and Partridge (2009) showed that about 50% of the U.S. workers could be uniquely identified at a census block level using only home and work locations coming from Longitudinal Employer-Household Dynamics (LEHD) [7]. This study revealed that the median size of the anonymity set of the workers in the U.S. at the Census block level is one. An anonymity set is a set of individuals that share the same attributes and cannot be distinguished from each other by the available information. Chow and Mokbel (2011) found that by having the paths of all users and knowing that a particular device was in certain places at certain times, they could identify the device's trajectory [8]. This statement will be used for the deduplication validation in the paper. In other words, for two devices to be identical, they must be observed at the same place at any time.

In Zang, Hui, and Jean Bolot (2011) study, CDR data were used to identify mobile device owners by analyzing their top N locations based on how often they appeared across different geographical levels such as sectors, cells, and zip codes [9]. A device whose top locations are fewer would be more challenging to identify. This study analyzed the top one, two, and three locations regarding the frequency of observations. Based on this research, more than half of the users could be uniquely identified by having their top three locations at the cell and sector levels. In addition, the top two locations were analyzed while they were interchangeably observed, as the users might make more calls from their work location than from their home location in one month while vice versa in another.

Instead of analyzing the top N locations, De Montjoye et al. (2013) investigated the number of random points needed to identify an individual mobility trace [10]. They evaluated the call data for 1.5 million users and found that about 95% of the people could be uniquely identified by four spatiotemporal observations from each device. Furthermore, in the re-identification of the individuals, both the spatial and temporal resolution of the devices' location observations are critical.

Human mobility patterns are highly predictable. Song et al. (2010) studied users' trajectories and noted that people tend to spend most of their time in a few locations [11]. According to the researchers, there is a potential for 93% predictability of average mobility, which does not vary much by population. In another study, Gonzalez et al. (2008) showed that the human mobility of individuals is consistent in both spatial and temporal domains and that people tend to return to their preferred locations regularly [12]. We believe the main research gap in this field is the lack of studies evaluating duplicate devices among mobile device location data provided by different data vendors. As mentioned above, predictability of the human mobility pattern can lead to determine if sightings of two devices belong to the same person. Based on the reviewed literature, a device deduplication methodology is proposed and validated.

**METHODOLOGY**

Based on the literature review, a deduplication algorithm utilizing both demographic information, i.e., the home location of every device, and their travel pattern defined as the location visited by the device in a timespan is proposed and validated. The deduplication algorithm considers devices with the same home location and in-order top five most visited locations in a month to be in the same k-anonymity, and flags them as duplicate ones. This means these devices represent the same user but with different device IDs and need further treatment. The home location identification algorithm and how the value "five" for the top visited location is derived and validated are described in the following sections.

*Data*

The dataset used in this study is the mobile device location data (MDLD) collected by multiple leading data vendors. This dataset contains the spatial and temporal information of several users, including a random hashed device identifier, the latitude and longitude of location points in decimal version, the time that the location of the user has been collected as a timestamp, the accuracy of the sightings as meters reported by the data provider, and the Coordinated Universal Time (UTC) offset that relates the UTC of each sighting to their local time. Table 1 shows a sample of the mobile device location data. Due to privacy concerns, noise has been applied to all the entries shown in Table 1. This table is just an imaginary dataset for the purpose of showing the information provided by MDLD.

Table 1. A sample of MDLD.

| Device-ID | UTC timestamp | Latitude | Longitude | Accuracy | UTC offset |
|---|---|---|---|---|---|
| Sfbcx-223da | 1578010770 | 38.9924 | -76.9293 | 2 | -14400 |
| Sfbcx-223da | 1578010775 | 38.9802 | -76.9190 | 5 | -14400 |
| Sfbcx-223da | 1578010778 | 38.9605 | -76.9201 | 3 | -14400 |
| Rjckf-2421s | 1578010500 | 38.7069 | -76.8985 | 11 | -14400 |

To illustrate the coverage of the utilized dataset, a sampling rate at the county level is calculated. Sampling rate is defined as the number of devices with an imputed home location divided by the total population of the county. Figure 1 shows the mobile device location data sampling rate. 92% and 90% of the counties have a sampling rate of more than 5% and 10%, respectively. The numbers indicate the great value of MDLD for the analysis of travel patterns, compared with surveys in which the sampling rate is much lower than the sampling rate in MDLD used in this study. Furthermore, the utilized dataset is one month of MDLD provided by multiple data vendors, January 2020.

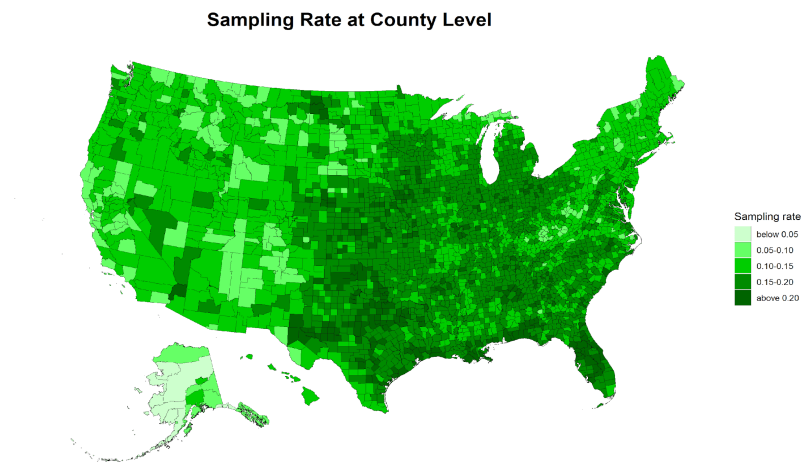

Figure 1. Sampling rate at the county level.

*Geographical Level of Study*

The geographical level of study for users' home locations and visited locations is a geo-hash level 7. Geo-hashes are unique identifiers of specific zones on the earth, and their width and height depend on the level of the certain geo-hash. Table 2 shows geo-hashes' size, from the largest to the smallest.

Table 2. Geo-hash width and height at different levels.

| Level of geo-hash | width × length |
|---|---|
| 1 | 5,009.4 km × 4,992.6 km |
| 2 | 1,252.3 km × 624.1 km |
| 3 | 156.5 km × 156 km |
| 4 | 39.1 km × 19.5 km |
| 5 | 4.9 km × 4.9 km |
| 6 | 1.2 km × 609.4 m |
| 7 | 152.9 m × 152.4 m |
| 8 | 38.2 m × 19 m |
| 9 | 4.8 m × 4.8 m |
| 10 | 1.2 m × 59.5 cm |
| 11 | 149 mm · 149 mm |
| 12 | 37.2 mm · 18.6 mm |

Figure 2 illustrates the level of study for home location imputation and visited locations. The blue rectangle is a geo-hash level 7 denoted by the unique identifier named "dqcmc4p" with a size of 152.9 m × 152.4 m. These sizes are only true at the equator and the actual sizes may differ depending on the latitude of the location.

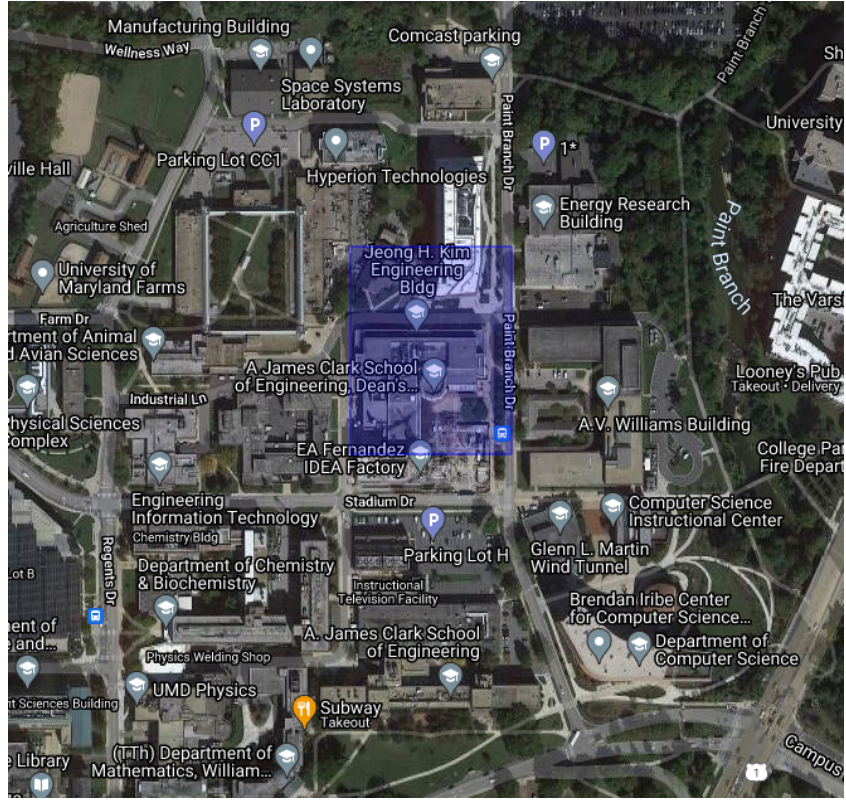

Figure 2. Illustration of a level-7 geo-hash.

*Home Location Identification*

The proposed algorithm identifies duplicate users based on their home locations and frequently visited places. Specifically, their home is considered where people spend most of their time. Thus, home location is the most significant feature in identifying duplicate devices. Since the MDLD does not contain anyground truth information of the device owners, the home locations need to be imputed from the location sightings. The study adopted the home location identification algorithm from the doctoral dissertation of one of the authors. [13].

Each device's home location is identified from the geo-hash zones the device visited during a month. The algorithm first identifies the home and workplace at level-6 geo-hash due to the uncertainty of location sightings and the minor movements around the home. Next, it selects the most frequently observed level-7 geo-hash within the identified level-6 geo-hash as a more precise representation of the imputed home location. The methodology is as follows: At first, the sightings of a device are aggregated at geo-hash level 6 zones, and geo-hashes that meet the following criteria are kept:

- Observed for at least three days in the month.
- Observed for more than half of the total days that the device was observed in the month.
- Observed at two or more distinct hours on average from all the days the candidate location was observed.

The remaining geo-hashes are sorted based on the number of observed days in a month, the average daily number of observed hours in observed days, and the average number of hourly sightings in observed hours. Accordingly, the top three are picked. Next, the top three remaining geo-hashes are sorted by the observed number of nights, the average daily number of observed nighttime hours, and the average number of hourly sightings during nighttime. The top geo-hash level 6 is identified as the home location, since people tend to spend most of their nighttime at home. Finally, these two steps are repeated on all the level-7 geo-hashes in the identified home level-6 geo-hash, and the top-ranked level-7 geo-hash is selected as a more precise representation of the home location. It is worth noting that the nighttime hour is chosen to be 9 pm to 6 am based on the American Time Use Survey (ATUS) that shows nearly 80% of the population that work full-time or part-time visit the home location at this time.

The home location identification also serves as an additional quality filter on the LBS data. It is widely observed that the LBS data from different devices can have different data qualities; thus, some devices cannot be utilized due to insufficient data. For example, some devices have less than 10 location sightings for the entire month, whose travel patterns cannot be confidently reproduced. After the home location identification process, only devices with a minimum quality will remain in the dataset.

*Device deduplication*

Based on both demographic information - in this case, the imputed home location of devices- and travel movement pattern- the top N locations visited during a month- a deduplication algorithm is developed to identify different device IDs that represent the same user in the integrated dataset coming from several data vendors. In this study, devices that have the exact home location and top-five most-visited locations are considered in the same k-anonymity and, as such, will be flagged as duplicate devices. These devices represent the same user but have different device identifiers.

Geo-hashes with at least one sighting over a month are considered as visited locations of a device. To determine the top locations visited by a device, all the visited geo-hashes are sorted based on the number of unique hours and the number of sightings during a month. A unique hour is a time interval of one hour during which a specific device was observed. For example, two observations at 11:45 a.m. and 11:12 a.m. on January 4 and January 6, 2020, are considered two unique hours during a month. Next, the top N geo-hashes are chosen as the most-visited locations of a device for the deduplication process. Devices that do not have top N geo-hashes, meaning they are observed in N-1 or fewer geo-hashes during the month, are removed since the information about their trajectories is insufficient.

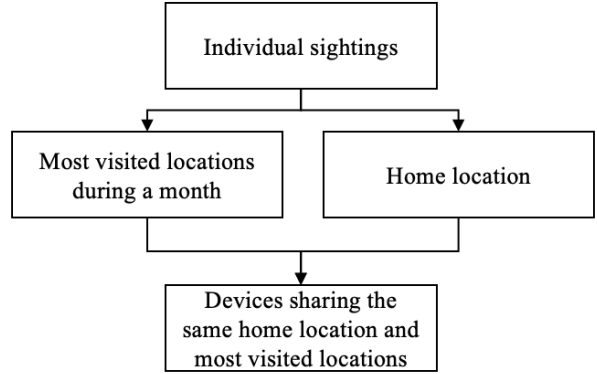

Figure 4. Basic diagram of the steps to identify duplicate devices. First, sightings of each device is processed to identify their home location as well as their top most visited locations during a month. Devices sharing the same home location and most visited locations are recognized as duplicates.

**RESULTS**

Based on the framework described in the methodology section, a user deduplication method is developed, calibrated, and validated. Table 3 summarizes the minimum, average, and maximum values of anonymity group sizes for devices in the dataset. As the number of the top-visited locations increases, fewer devices are associated with the same anonymity group.

Table 3. K-anonymity size statistics for devices having the exact home location.

| Number of the top-visited locations (N) | Min | Mean | Max |
|---|---|---|---|
| 1 | 1 | 4.23 | 20514 |
| 2 | 1 | 1.82 | 323 |
| 3 | 1 | 1.48 | 164 |
| 4 | 1 | 1.34 | 72 |
| 5 | 1 | 1.25 | 52 |
| 6 | 1 | 1.19 | 50 |
| 7 | 1 | 1.15 | 14 |
| 8 | 1 | 1.12 | 7 |

*Kabiri, Darzi, Saleh Namadi, Pan, Zhao, Sun, Yang, and Ashoori*

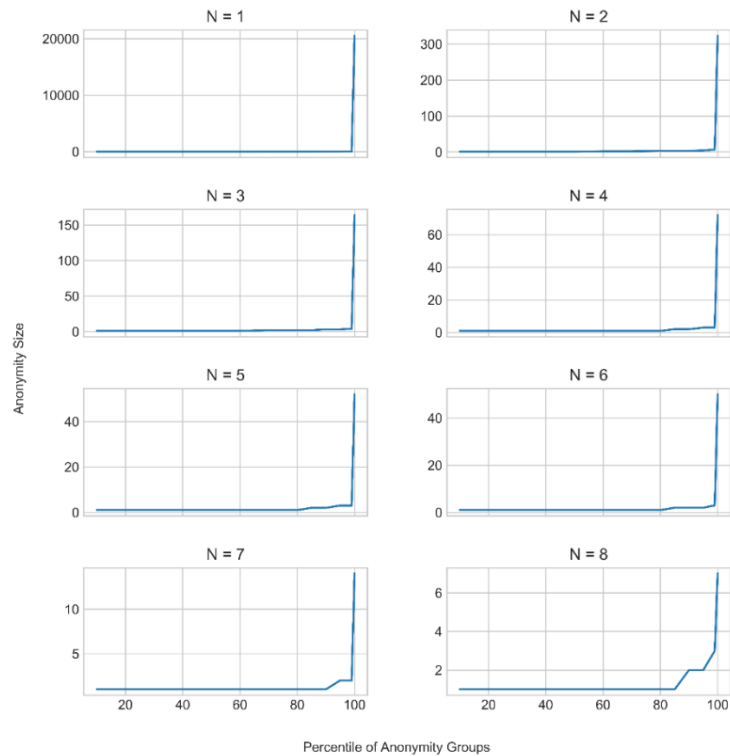

Figure 4. Anonymity size variation with different numbers for most-visited locations.

Figure 4 shows the percentile of anonymity sizes having different numbers of locations (N) as the most-visited locations. For example, with N equal to five, more than 80% of anonymity groups contain only a single device, and less than 20% have two or more devices in the same anonymity group. Previously, we noted that the objective is to consider all devices belonging to the same anonymity group representing the same individual since they share the top N most-visited locations and home locations. The next step is to determine a reasonable value for N. Based on Table 3 and Figure 4, as the value of N increases, fewer devices share the same most-visited locations, and consequently, fewer devices are identified as duplicates. A validation method is conducted to ensure the proper value of N is selected.

Sharing the top one (N=1) or two (N=2) locations is not sufficient to identify duplicated devices since these two locations are most likely to be the home and work location of the devices, and it is possible to have two devices living and working together. Thus, more locations are needed to be considered as duplicate identifiers. Figure 5 shows the number of devices with at least one duplicate with varying values of N. Higher values can lead to failing to detect duplicate devices because even tiny changes in the number of observed hours can lead to failing to detect duplicate devices. This is because the data of different data vendors may report different numbers of sightings of a device as different applications capture them.

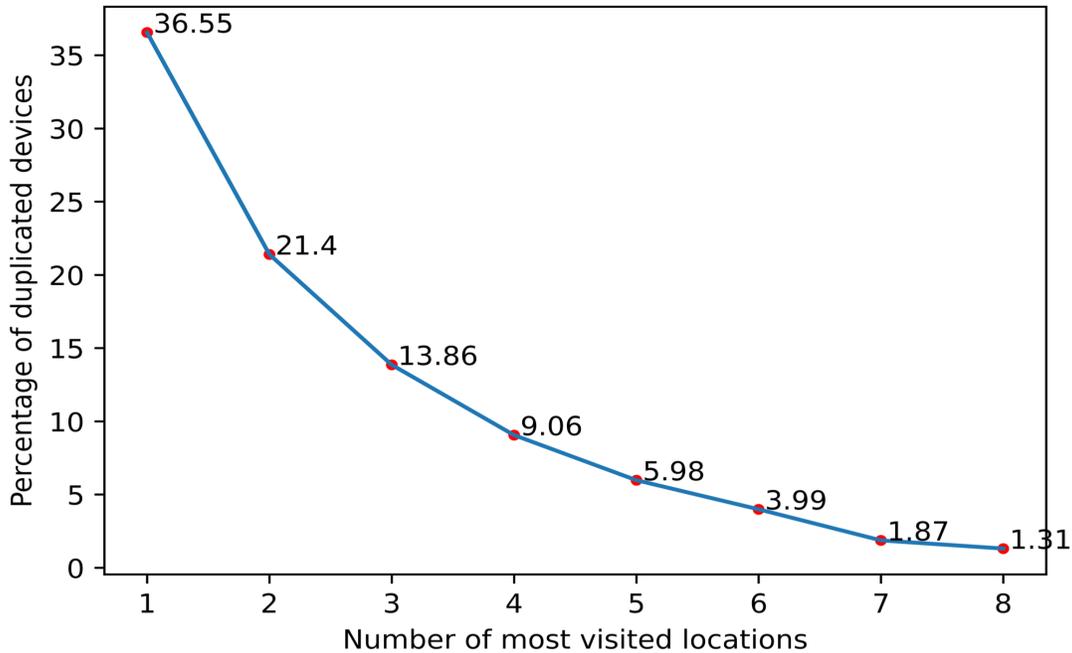

Figure 5. The number of devices having one or more duplicates in the dataset.

If two devices are duplicates, meaning they represent the same user in the dataset, they must simultaneously be in the same location. This fact is the baseline for validating the duplicate device detection framework. For the validation purpose, after applying the proposed methodology of device deduplication, 100,000 duplicated paired devices are evaluated in terms of spatial and temporal information of sightings in 10 days. Their trajectories are evaluated to see if they are at the same location simultaneously. Since different data vendors might report the sightings of the same user at different hours of a day, only the sightings of the common hours are compared. Common hours are those one-hour intervals in which both devices had sightings in the integrated dataset. For instance, if two devices appeared simultaneously at 4 a.m. on January 1, 2020, they must have appeared in the same level- 7 geo-hash.

Figure 6 shows the number of paired devices in the random sample data that had at least one common hour in the first ten days of January. As the number of most visited locations increases, the number of devices having sightings in the dataset at the same hour-intervals increases. For the values greater than five, the number of paired devices having common hours does not increase significantly.

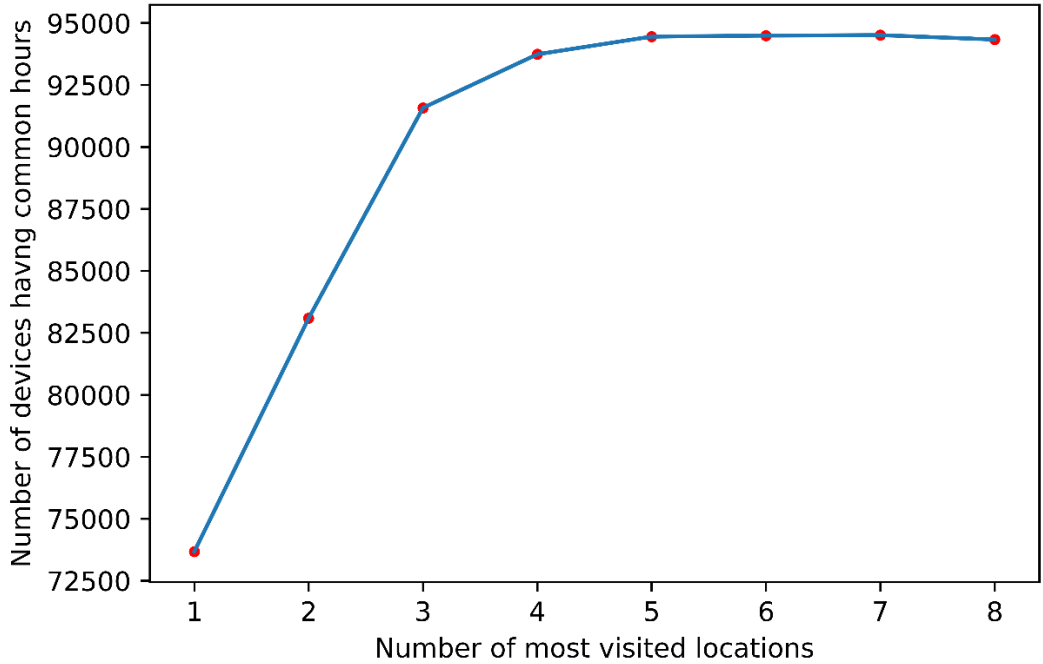

Figure 6. The Number of devices having common hours in the sample dataset.

Figure 7 shows the average percentage of common hours of paired devices observed in the same level-7 geo-hash. Same as the previous trend, when the value of N changes from one to five, the percentage increases a lot, but for the higher values, the percentage is almost around 99.5% and does not change a lot. This number is a good indicator of being in the same location simultaneously, and using a higher value increases the risk of failing to identify two duplicated devices.

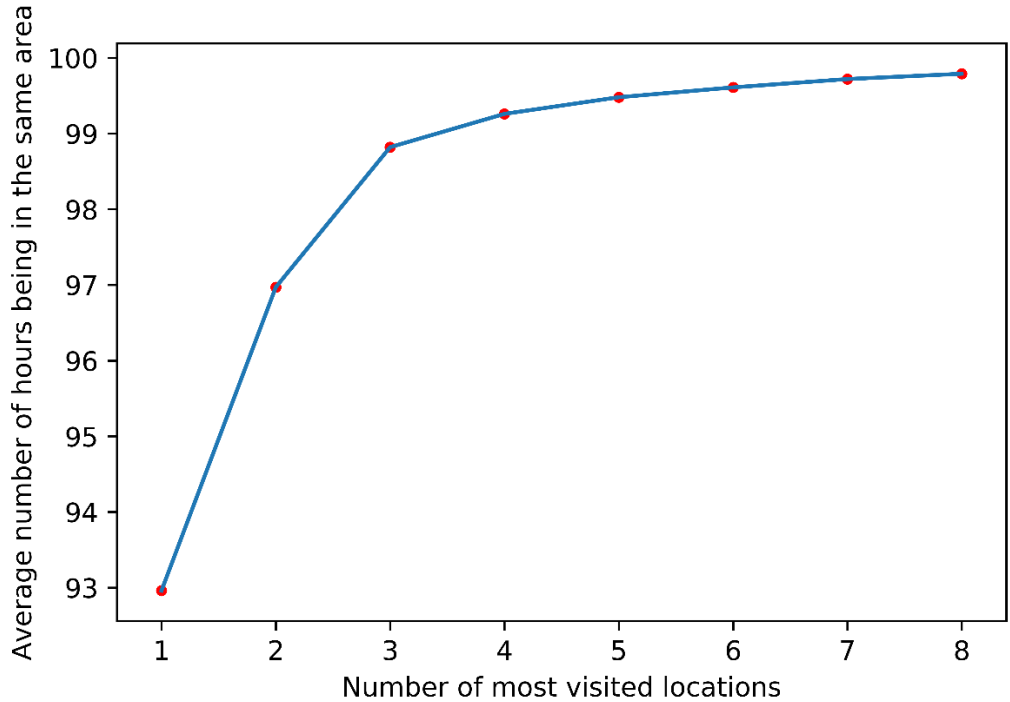

Figure 7. The average number of hours observed in the same level-7 geo-hashes.

Based on the previous discussions, a value of "5" for the number of most-visited locations of a device, along with sharing the same home location, is a proxy for the duplicate identifier. Finally, when two devices are labeled as duplicate devices, all their sightings are integrated, and the same device ID is assigned to them. In this way, a more solid trajectory of the user is presented in the final dataset. As an illustration, Figures 7-a and 7-b show the centroid of level-7 geo-hashes that two devices with different user identifiers in two datasets visited during the first ten days of January 2020. These two devices are identified as duplicates, and their sightings were merged. In this case, 100% of the time that both devices reported their location information were at the same level-7 geo-hashes.

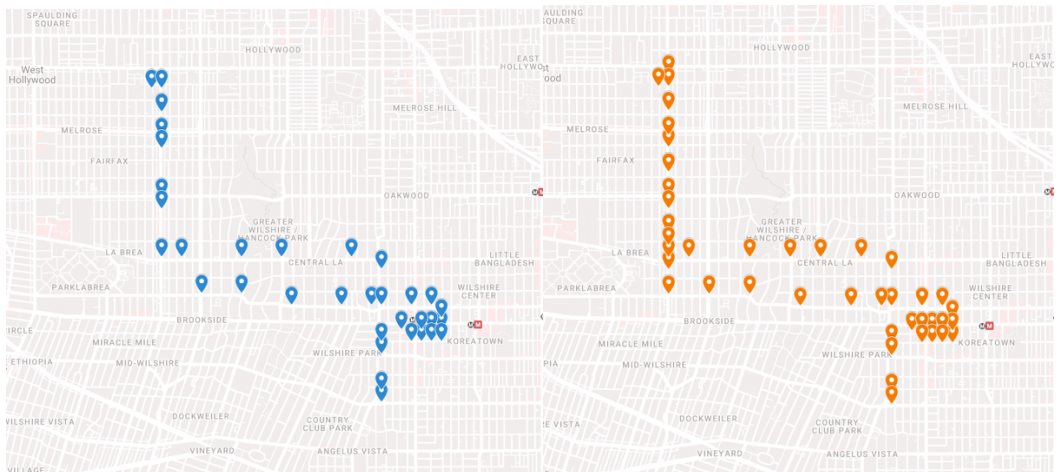

Figure 7-a.          Figure 7-b.

## CONCLUSION AND DISCUSSION

This study presents a user deduplication framework utilizing mobile device location data (MDLD) collected by several data vendors. The MDLD from various data vendors are integrated, and several data cleansing steps are carried out to get a solid raw dataset. A deduplication algorithm is developed to identify duplicate devices in the integrated dataset, and the sightings of such devices are merged to avoid overrepresented users. In this framework, user sightings at a level-7 geo-hash are examined spatially and temporally. Devices with the same home location and the top five most visited locations during a month represented the same user. This framework identified nearly 6% of the data subjects as duplicates and consolidated their sightings. In addition, the results of the study are validated to see if duplicate devices are observed simultaneously in the same location. As shown by a 100,000-sample of identified duplicate data subjects, almost 99.5% of them where at the same location at the same time throughout the study period. The framework proposed in this study has practical implications. In 2020, FHWA launched the Next-Generation National Household Travel Survey (NextGen NHTS) to provide a more continuous travel behavior of the U.S. population. NextGen NHTS provides an origin-destination (OD) data program that produces passenger travel OD tables at different geographical levels derived from MDLD. As part of the data preprocessing step of this project, the proposed user deduplication framework is applied [14].

Moreover, this study shows that people's mobility pattern can be a strong identifier of them. Findings indicate that one should utilize these data carefully in order to protect the privacy of individuals, such as revealing the results at aggregate levels only. As the future direction, an unsupervised machine learning approach can be implemented on the users' location data to identify duplicate users using activity location clusters throughout the study period. It is worth mentioning that concerning limitation of the clustering approach will be it results in a much more extensive computation both in terms of cost and time when analyzing the travel patterns of millions of users at the national level and it needs to be considered when developing such a framework.

## ACKNOWLEDGEMENTS

We would like to thank and acknowledge our partners and data sources in this effort: (1) partial financial support from the U.S. Department of Transportation's Federal Highway Administration; and (2) Amazon Web Services and its Senior Solutions Architects, Jianjun Xu and Greg Grieff, for providing cloud computing and technical support.

## CONFLICT OF INTEREST

The authors declare that they have no conflict of interest.

## AUTHOR CONTRIBUTION STATEMENT

The authors confirm their contribution to the paper as follows: study conception and design: A.K., A.D.; data collection: A.K., Y.P., A.D.; analysis and interpretation of results: A.K., S.S.N., Y.P., A.D.; draft manuscript preparation: A.K., Y.P., A.D., C.X., S.S.N., M.A., G.Z., Q.S., M.Y.

**REFERENCES**

1. Chankaew, N., Sumalee, A., Treerapot, S., Threepak, T., Ho, H. W., & Lam, W. H. (2018). Freight traffic analytics from national truck GPS data in Thailand. Transportation research procedia, 34, 123-130.

2. Chen, C., Ma, J., Susilo, Y., Liu, Y., & Wang, M.. The promises of big data and small data for travel behavior (aka human mobility) analysis. Transportation research part C: emerging technologies, 2016. 68, 285-299.

3. Yang, M. (2020). Multimodal Travel Mode Imputation Based on Passively Collected Mobile Device Location Data (Master's Thesis, University of Maryland, College Park).

4. Sweeney, L. (2000). Uniqueness of simple demographics in the US Population, in LIDAP-WP4. http://privacy. cs. cmu. edu/dataprivacy/papers/LIDAP-WP4abstract. html.

5. Golle, P. (2006). Revisiting the uniqueness of simple demographics in the US population. Proceedings of the 5th ACM Workshop on Privacy in Electronic Society.

6. Trestian, I. et al. (2009) Measuring serendipity: connecting people, locations and interests in a mobile 3G network. Proceedings of the 9th ACM SIGCOMM conference on Internet measurement.

7. Golle, P., & Partridge, K. (2009). On the anonymity of home/work location pairs. International Conference on Pervasive Computing, Springer, Berlin, Heidelberg.

8. Chow, C.Y., & Mokbel, M.F. (2011). Trajectory privacy in location-based services and data publication. ACM Sigkdd Explorations Newsletter, 13(1), 19-29.

9. Zang, H., & Bolot, J. (2011). Anonymization of location data does not work: A large-scale measurement study. Proceedings of the 17th annual international conference on Mobile computing and networking.

10. De Montjoye, Y. A., Hidalgo, C. A., Verleysen, M., & Blondel, V. D. (2013). Unique in the crowd: The privacy bounds of human mobility. Scientific reports, 3(1), 1-5.

11. C. Song, Z. Qu, N. Blumm, and A.-L. Barabasi. Limits of predictability in human mobility. Science, 327(5968):1018{1021, 2010.

12. Gonzalez, M. C., Hidalgo, C. A., & Barabasi, A. L. (2008). Understanding individual human mobility patterns. nature, 453(7196), 779-782.

13. Pan, Y. (2021). National-Level Origin-Destination Estimation Based on Passively Collected Location Data and Machine Learning Methods (Doctoral dissertation). University of Maryland. https://doi.org/10.13016/fnh1-jewx.

14. Federal Highway Administration. (2020). 2020 NextGen NHTS National Passenger OD Data, U.S. Department of Transportation, Washington, DC. Available online: https://nhts.ornl.gov/od/.